# Mathematical Modeling and Computer Simulation of Needle Insertion into Soft Tissue

Adam Wittek[1*], George Bourantas[1], Benjamin F. Zwick[1], Grand Joldes[1], Lionel Esteban[2], Karol Miller[1]

[1]*Intelligent Systems for Medicine Laboratory, The University of Western Australia, Perth 6009, Western Australia, Australia*
[2]*Commonwealth Science and Industry Research Organization CSIRO, Medical XCT Facility, Kensington 6151, Western Australia, Australia*

**Abstract:** In this study we present a kinematic approach for modeling needle insertion into soft tissues. The kinematic approach allows the presentation of the problem as Dirichlet-type (i.e. driven by enforced motion of boundaries) and therefore weakly sensitive to unknown properties of the tissues and needle-tissue interaction. The parameters used in the kinematic approach are straightforward to determine from images. Our method uses Meshless Total Lagrangian Explicit Dynamics (MTLED) method to compute soft tissue deformations. The proposed scheme was validated against experiments of needle insertion into silicone gel samples. We also present a simulation of needle insertion into the brain demonstrating the method's insensitivity to assumed mechanical properties of tissue.

## 1. INTRODUCTION

Needles are frequently used in various medical procedures such as drug delivery (1), tissue biopsy (2, 3), blood sampling (4), anaesthesia (5) and radiation cancer treatment among others (6-8). Accurate needle tip placement is important, since most procedures rely on accurate targeting for an effective outcome. A missed target may prevent delivery of a therapeutic agent and worse, may damage neighboring structures. Targeting in needle insertion is seemingly simple: one aims a rigid needle at a fixed point whose position is known *a priori* from medical images. In reality, the target moves due to tissue deformations caused by the needle and overall organ motion.

Current methods for addressing these coupled phenomena (9-12) use linear models of target motion which, while fast, are inaccurate. This leads to a need for continuous image guidance. Even with the latest intraoperative 3D imaging technology, such tracking is limited in spatial and temporal resolution, expensive, and thereby not always feasible (13). Currently available imaging technologies for direct tracking of the intraoperative motion of anatomical targets exhibit important limitations: *i)* ionizing radiation exposure in X-ray and Computed

Tomography (CT) (14, 15); *ii*) noisy images and inability to penetrate the bone/skull in Ultrasound (16); *iii*) slow acquisition (order of at least several minutes) in Magnetic Resonance Imaging (MRI) (17-19); *iv*) ability to track only the surgically exposed organ surface in navigation systems using cameras (20).

Instead of focusing on better target imaging, we concentrate on improving predictive models. We use needle motion as an input to compute deformations within the organ and predict the motion of a target during surgery. Previous approaches for such prediction relied on finite element discretization (21-24) and tended to use the assumptions of linear elasticity (that simplify the body organs as continua with linear stress-strain relationship undergoing infinitesimally small deformations (22, 25-28)). There are major obstacles associated with such approaches:

- Oversimplifying and unrealistic assumptions: **(a)** Surgical procedures induce large strains (as high as 80% at needle tip, see e.g. (29)) and discontinuities (due to needle insertion) in the body organs. **(b)** The vast body of experimental evidence indicates that soft tissues exhibit nonlinear stress-strain behavior (30).
- Time consuming generation of computational grids (finite element meshes): Robust and accurate computation of organ deformation requires good quality finite element meshes. For organs with complex geometry, such as the brain, generation of such meshes is time consuming even if state-of-the-art software for anatomic mesh generation is used (31-33).

The problems indicated above motivate our research into mathematical modeling and computer simulation of needle insertion. Our overall goal is to create methods for robust patient-specific simulations that in the future could be used in needle guidance systems to refine pre-operative surgical plans leading to more accurate and faster targeting and better outcomes.

For the last four decades, Finite Element Analysis (FEA) has been the method of choice in computational biomechanics. Nevertheless, the conventional approach to compute soft tissue deformation relies on linear finite element algorithms that assume infinitesimal deformations (22, 34). However, modeling of soft tissue organs for surgical simulation and image-guided surgery is a non-linear problem of continuum mechanics which involves large deformations and large strains with geometric and material non-linearities (35-37), clearly incompatible with the assumption of infinitesimality of deformations. Co-rotational finite elements (38) were proposed to allow close-to-real time computation of deformations. However, this formulation assumes small strains, assumption clearly not satisfied in many clinically relevant scenarios. Another difficulty with using the Finite Element Method for patient-specific applications arises from the common practice of using 4-noded tetrahedral (i.e. linear) finite elements. These elements exhibit volumetric locking and should not be used for almost incompressible materials such as soft tissues (39-42). Parabolic (10-noded) tetrahedron is appropriate but computationally inefficient (43). 8-noded hexahedra are preferable, but efficient generation of hexahedral meshes for complicated geometries, despite enormous research effort (44), still awaits a satisfactory solution (32). Using FEM for simulating needle insertion is even more problematic due to very large strains at the needle tip - 80% strains were seen close to the tip of a needle inserted into swine brain (29) - leading to element distortion and necessity to remesh.

In this study, we develop and solve our models using the Meshless Total Lagrangian Explicit Dynamics (MTLED) algorithm that accounts for very large deformations and strains, nonlinear stress-strain behavior of soft tissues, and utilizes a trivial to construct unstructured cloud of points as the computational grid (31). Algorithms such as this overcome costly computational grid generation and are very effective for problems involving large strains and surgical tool insertion (31, 45-47).

The key innovative element of our approach is the focus on patient-specific modeling and simulation without relying on very difficult to measure patient-specific properties of tissues (48-50), patient-specific parameters of needle-tissue contact models (51, 52), and tissue damage models (53). We propose a kinematic approach to modeling the tissue-needle interaction, with parameters of the model identifiable from images.

The paper is organized as follows: the advantages of MTLED algorithm (31) are succinctly described in Section 2.1; our novel kinematics-based approach to model needle insertion is presented in Section 2.2; verification of our methods is presented Sections 2.3 and 3.1, and experimental evaluation is described in Sections 2.4 and 3.2. We highlight the applicability of the proposed method to analysis of continua with complex geometry in Sections 2.5 and 3.3 Discussion of our results is in Section 4.

## 2. MATERIALS AND METHODS

## 2.1 Computational Solid Mechanics Framework: Meshless Total Lagrangian Explicit Dynamics (MTLED)

In patient-specific applications, where compatibility with clinical workflow is of essence, restrictions on time and effort required to generate spatial discretization limit the use of finite element (FE) method for solving nonlinear partial differential equations governing the deformations of soft tissues (32). Instead, we use the Meshless Total Lagrangian Explicit Dynamics (MTLED) algorithm capable of overcoming the FE limitations through the use of an unstructured cloud of nodes (instead of interconnected elements) to discretize the geometry. MTLED was first proposed in (54) and comprehensively developed and described in (31). Here we only restate the major advantages of the algorithm.

In MTLED, nodal generation is automatic since the nodes can be arranged/distributed in almost arbitrary way (55). Another important advantage of meshless discretization over the mesh of interconnected elements is the ability to deal with extremely large deformations and boundary changes that occur during surgical procedures such as needle insertion, retraction, resection and tissue removal.

MTLED is formulated in Total Lagrangian framework which allows all quantities to be computed with respect to the initial configuration and consequently all spatial derivatives used in the algorithm can be precomputed, resulting in substantial savings in computational effort (39). The method involves three stages: pre-processing, solution and post-processing. In the pre-processing step, all the geometry and material properties are defined. The spatial domain is represented by nodes (displacements and forces are calculated on the nodes; mass is assigned to them) and integration points (where stresses and strains are calculated). Both nodes and integration points exist as particles in the geometry with no connection to each other before support domains and shape functions are created. In the solution step, the displacement field is computed using an explicit time integration scheme. The critical time step, needed for the conditionally stable explicit scheme, is computed during the simulation (56). There is no need to solve a linear system, therefore the method is easy to apply and parallelize. The post-processing step involves computation of derived quantities (apart from displacement field) such as strain and forces.

## 2.2 Modeling needle insertion – kinematic approach

*2.2.1 Kinematic approach*

To avoid the need for patient-specific material properties and needle-tissue interaction models, we propose a novel kinematic modeling approach following the ideas we previously outlined in (57). The proposed method directly links the deformation of the tissue adjacent to the needle tip and along needle shaft to the needle motion.

Following the experimental literature on needle insertion into soft tissues (58) two phases are distinguished in our kinematic approach: (*i*) **indentation**, where the organ surface deforms as a result of contact interactions with the needle tip; and (*ii*) **tissue penetration** by the needle that follows the puncture of the organ surface by the needle tip. During the tissue penetration phase, the tissue is in contact with both the needle tip and the needle shaft.

- *Indentation***:** During indentation, only a small area on the organ surface is in contact with the needle (we represent this area by a subset of nodes located on the organ surface). The displacement of these nodes equals the known (imposed) needle tip displacement. We monitor strain in the needle insertion area, and, when the strain exceeds a threshold value (referred to as puncture strain $\varepsilon_p$), the needle punctures the organ surface and the penetration phase starts.

- *Tissue Penetration***:** During penetration, we define the nodes located close to the needle shaft, and we displace them by a fraction (referred to as the deformation coefficient $C_D$) of the known (imposed) displacement of the needle. This approach removes the need for a patient-specific needle-tissue mechanical interaction model.

Thus, our method for needle insertion modeling has only two parameters ($\varepsilon_p$ and $C_D$). Both can be determined from images of the continuum/body organ undergoing needle insertion as explained in section *2.2.3 Parameters for the kinematic approach*.

*2.2.2 Material model*

As we demonstrate in the following sections, our modeling method allows accurate computation of tissue displacements without knowledge of material properties of the tissue. Nevertheless, for method verification purposes we identified mechanical properties of gels used in our experiments.

We conducted our experiments using Sylgard 527 silicone gel that has been reported in the literature as exhibiting the mechanical behavior similar to the brain tissue (59, 60) and is regarded as scientifically accepted brain tissue surrogate (61-63) . It should be noted here that the bio-fidelity of different materials in representing the brain tissue mechanical responses is a subject of extensive research (23). However, such research is beyond the scope of this study.

To determine the mechanical response of the gel sample, and to describe the non-linear stress-strain mechanical response of nearly incompressible materials we use an Ogden material model. Our previous research (60) has indicated that Sylgard 527 material behavior can be represented using Ogden model:

$$W = \frac{2\mu}{a^2}\left(J^{-\frac{a}{3}}\lambda_1^a + J^{-\frac{a}{3}}\lambda_2^a + J^{-\frac{a}{3}}\lambda_3^a - 3\right) + \frac{1}{D}(J-1)^2 \qquad (1)$$

where $\lambda_1, \lambda_2, \lambda_3$ are the principal stretches; $a$, $\mu$ and $D$ are material constants. We determined the parameters from the semi-confined compression experiments (64, 65), as shown in Fig. 1. The identified parameters are $\alpha = -1.3$, $\mu = 722$ Pa, and $D = 5.57738 \times 10^{-5}$ Pa$^{-1}$ (Poisson's ratio $\nu = 0.49$). It cannot be stressed enough that we require accurate material description solely for method verification. In practical simulations, patient specific material constants are not needed. Only kinematic parameters described below are needed.

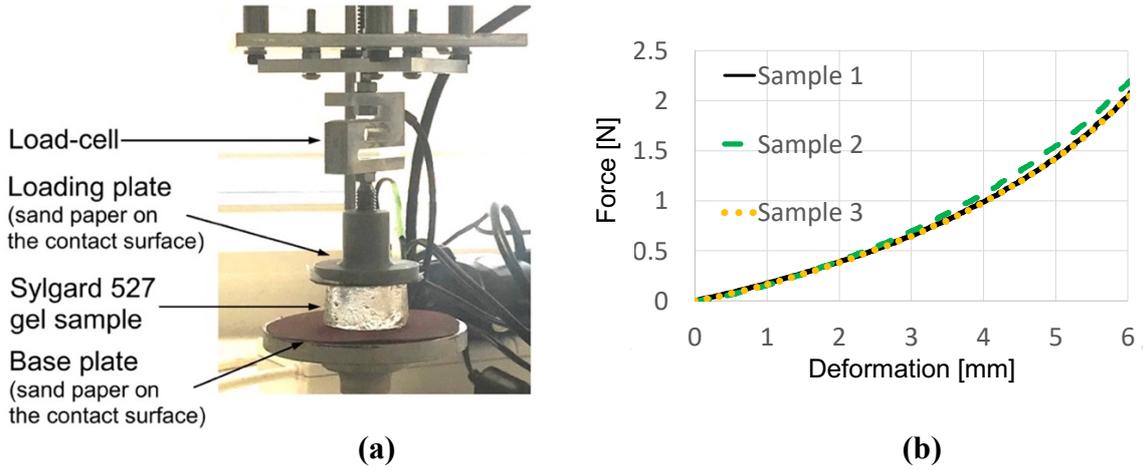

(a)                  (b)

**Fig. 1** (**a**) Semi-confined compression of a cylindrical sample (diameter of 30 mm, height of 17 mm) of Sylgard 527 gel for determining the gel material parameters. Technical specification of the apparatus is available at http://isml.ecm.uwa.edu.au/ISML/. We used Bestech KD40S-5N tension-compression load-cell with 5N force range (www.bestech.com.au). (**b**) Example of force measured in the experiments. We conducted the experiments for three gel samples from a given batch of gel.

*2.2.3 Parameters for the kinematic approach for needle insertion modeling*

Our algorithm for modeling needle–tissue interactions uses two parameters that can be determined from the images of the continuum undergoing deformation due to needle insertion: 1) Puncture strain $\varepsilon_p$ (when strain exceeds this value penetration initiates) and 2) Deformation coefficient $C_D$ (that links the displacement of the material adjacent to the needle with the displacement of the needle — see Fig. 2 for the geometry of needle used in this study). To illustrate this, we determine these two parameters by conducting needle insertion into the Sylgard 527 silicone gel samples and recording the sample deformation using X-ray C-arm General Electric 9900 apparatus. The experiments (Fig. 3) were conducted at The University of Western Australia Clinical Training and Evaluation Centre CTEC. For the gel samples used in the present study, we measured that gel adjacent to the needle moves/deforms (Sylgard 527 gel tends to firmly stick/attach to smooth surfaces such as surgical needle shafts) by ~40% of the distance travelled by the needle tip (therefore $C_D$=0.4).

When analyzing the gel sample deformations, we did not observe any spring-back that can be associated with the puncture. Furthermore, there was no visible instantaneous drop in the force acting on the needle that following the puncture (see Fig. 10). Therefore, puncture strain $\varepsilon_p$ was designated an arbitrarily small value of $\varepsilon_p = 10^{-5}$ to account for a very short indentation stage in our simulation.

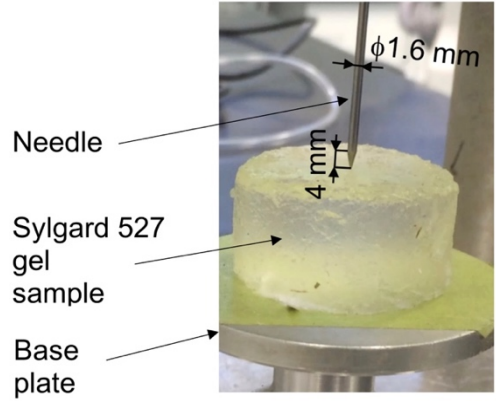

**Fig. 2** Geometry of the needle used in this study.

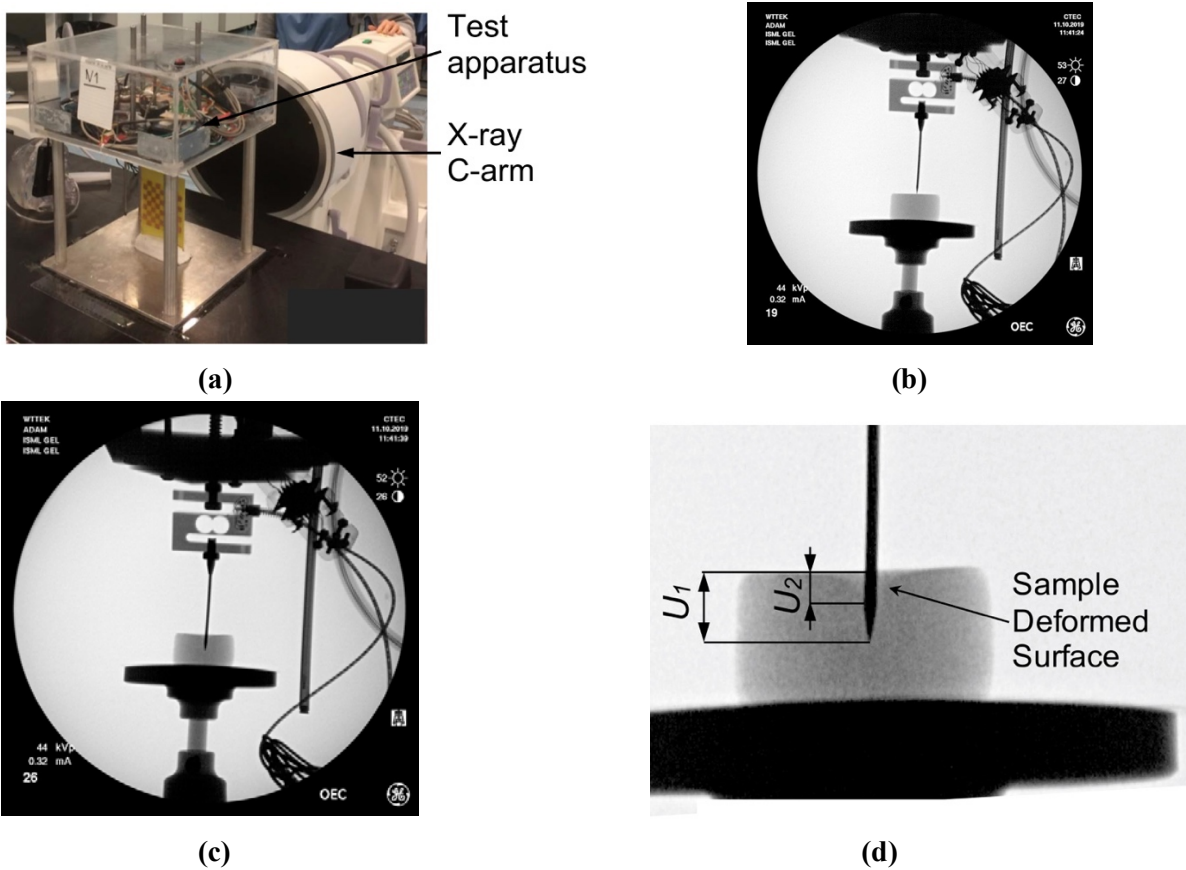

**Fig. 3** X-ray images of the cylindrical sample (diameter of 30 mm, height of 17 mm) of Sylgard 527 gel during needle insertion. The images were acquired using X-ray C-arm General Electric 9900 apparatus located at The University of Western Australia Clinical Training and Evaluation Centre CTEC. **(a)** Calibration of the X-ray apparatus image acquisition system and image distortion evaluation using the X-ray opaque chessboard-like calibration pattern (the pattern was machined from a printed circuit board PCB). **(b)** X-ray image of the sample at the start of needle insertion. **(c)** X-ray image of the sample after the needle insertion to the depth of 8 mm. **(d)** Locally enlarged X-ray image after the needle insertion to the depth of 8 mm. Gel deformation in the area adjacent to the needle is clearly visible. $U_1$ is the needle insertion depth (i.e. the needle tip displacement in relation to top sample surface) and $U_2$ is the maximum deflection of the sample surface along the needle shaft. The deformation coefficient (see section *3.1 Kinematic Approach*) $C_D = U_2/U_1$. For the experiment shown in this figure $C_D = U_2/U_1 \approx 0.4$.

## 2.3 Method verification

We verify our proposed modeling and simulation method by considering needle insertion into a homogeneous cylindrical sample (diameter 30 mm; height 17 mm; referred to as a *small sample*) made from Sylgard 527 gel (Fig. 4). This includes verification of our method convergence, demonstration of the method's ability to accurately compute the needle reaction force when the material properties of the sample are known, and demonstration of the computed displacement field independence of material properties.

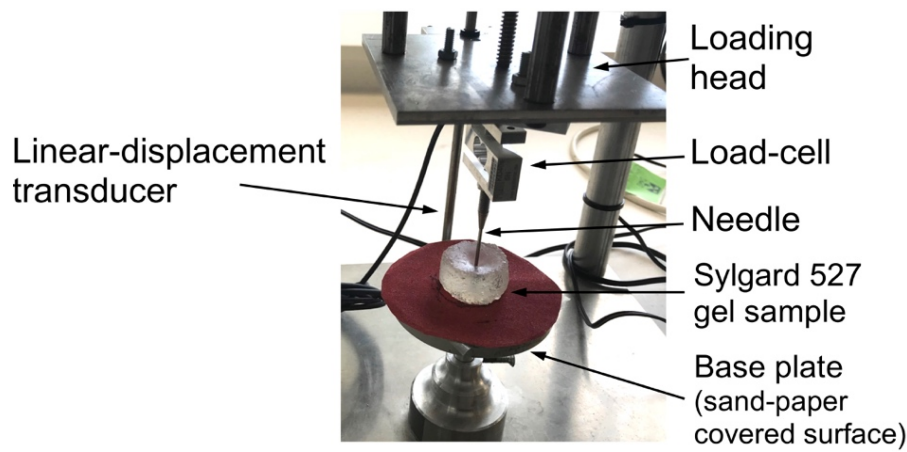

**Fig. 4** Experimental set-up for the needle insertion into silicone gel cylindrical (diameter of 30 mm and height of 17 mm) sample and needle force measurement. The needle insertion was conducted using the specialized apparatus we also applied in compression tests to determine Sylgard 527 gel material properties (see Fig. 1). The needle force was measured using Bestech KD40S-5N tension-compression load-cell with 5N force range (www.bestech.com.au).

*2.3.1 Verification of method convergence*

We modeled needle insertion (down to 15 mm) into the *small gel sample* using a successively denser nodal distribution to represent the spatial domain (the experimental set-up is shown in Fig. 4). The coarse grid used consists of 7,480 nodes and 39,145 integration cells (one integration point per integration cell); the moderate grid of 17,730 nodes and 96,038 integration cells, and the refined grid of 30,294 nodes and 166,843 integration cells (Fig. 5).

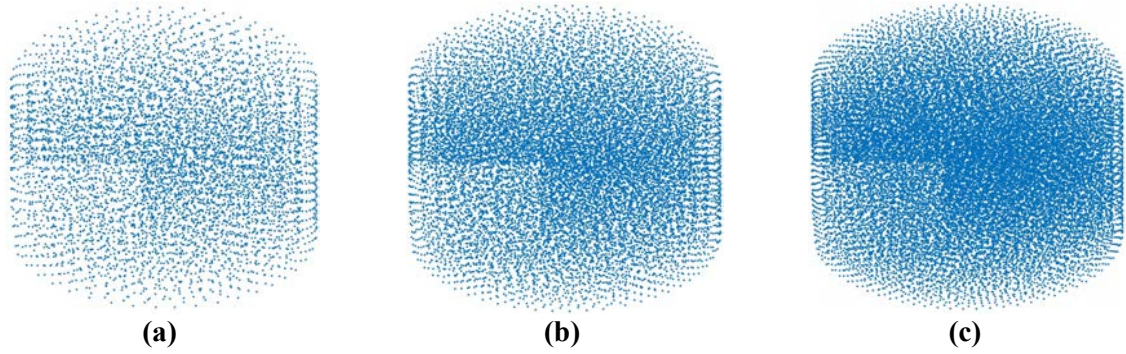

| (a) | (b) | (c) |

**Fig. 5** Meshless computational grids used when verifying convergence of the method for modeling of needle insertion proposed in this study. **(a)** Coarse grid (7,480 nodes); **(b)** Moderate grid (17,730 nodes); and **(c)** Refined grid (30,294 nodes). Convergence analysis was conducted through modeling of needle insertion into a cylindrical sample according to the experimental set-up shown in Fig. 4.

*2.3.2 Demonstration of the ability to compute needle reaction force*

Following the results of convergence analysis, we apply a computational grid consisting of 17,730 nodes and 96,038 integration cells (one integration point per cell) to simulate needle insertion (15 mm insertion depth) into a *small gel sample* and compare the computed and experimentally measured (the experimental set-up is shown in in Fig. 4) forces acting on the needle. The material behavior of Sylgard 527 gel is represented using the Ogden material model with the parameters determined in section 2.2.2 *Material model*: $a = -1.3$, $\mu = 722$ Pa and $D = 5.57738 \times 10^{-5} \ Pa^{-1}$.

*2.3.3 Demonstration of displacement field independence of material properties*

Following the results of our previous studies on computing deformations of soft tissue specimens subjected to uniaxial tension and compression (66) and predicting the brain deformations due to craniotomy (49, 50), we expect the deformations computed using our methods for needle insertion modeling to be independent of the selection/assumptions regarding the material model and stress parameter (shear modulus or Young's modulus). To demonstrate this, we apply the computational grid (17,730 nodes and 96,038 integration cells with one integration point per cell), used in section *2.3.2 Demonstration of the ability to compute needle reaction force*, to conduct simulation of needle insertion into a *small sample* (diameter of 30 mm and height of 17 mm) when varying the material properties (two orders of magnitude difference in the shear modulus) and material model (we used the Ogden and neo-Hookean models) as described in Table 1.

**Table 1** Parameters of the Ogden and neo-Hookean material models when evaluating the sensitivity of our method for needle insertion modeling to the material properties of the analyzed continuum. Note two orders of magnitude difference in stress parameter (shear modulus) between Material 2 and Material 3.

|  | $\alpha$ | $\mu\ (Pa)$ | $D\ (Pa^{-1})$ | Poisson's ratio |
|---|---|---|---|---|
| Material 1 (Ogden) | -1.3 | 72.0 | $5.57738 \times 10^{-4}$ | 0.49 |
| Material 2 (Ogden) | -1.3 | 722.0 | $5.57738 \times 10^{-5}$ | 0.49 |
| Material 3 (Ogden) | -1.3 | 7220.0 | $5.57738 \times 10^{-6}$ | 0.49 |
| Material 4 (neo-Hookean) | * | 1000.0 | — | 0.49 |

*Note: Neo-Hookean material model can be interpreted as a specific case of the Ogden model with $\alpha$=2.0.

### 2.4 Experimental evaluation of the method

To experimentally evaluate our modeling and simulation method we constructed a cylindrical sample (diameter of ~65 mm and total height of ~27 mm – see Fig. 5) with embedded steel beads whose displacements during needle insertion were tracked by 3D CT. Manufacturing of a sample required creation of three layers with 46 beads (diameter of 100 $\mu$m) inserted between bottom and middle layers, and middle and top layers (Fig. 6).

The bottom layer (Layer 1) has height ~18 mm, the middle one (Layer 2) ~9 mm, and the top one (Layer 3) ~7 mm. The material properties for each gel layer are given in Table 2. They were experimentally determined using semi-confined compression as described in section *2.2.2 Material model*.

Bead tracking was performed by acquiring a series of images of the gel sample in different stages of needle insertion (before the gel penetration by the needle, for the needle insertion depths of 5 mm and 15 mm) using a computed tomography Siemens SOMATOM AS medical scanner installed at Medical XCT Facility of Commonwealth Scientific and Industrial Research Organization (CSIRO) in Kensington, Western Australia (Fig. 7). XCT is a radiological imaging system first developed by Hounsfield (67). This non-destructive technique uses X-rays to obtain a three-dimensional data set of a sample by stacking contiguous cross-sectional two-dimensional images. In our experiments, we used an energy beam of 140kV/500mAs in helical mode acquisition every 0.10 mm (64 slices acquisition per 1 second rotation). A field of view of 71.86 mm × 71.86 mm was selected to oversee the whole gel sample together with the needle tip. This resulted in a voxel size of 0.16 mm × 0.16 mm × 0.10 mm. Each CT transversal (in X-Z plane) image (512 × 512 pixels) was reconstructed using Siemens algorithm (H70h) that enhances the sharpness of the images from edge detection density contrast.

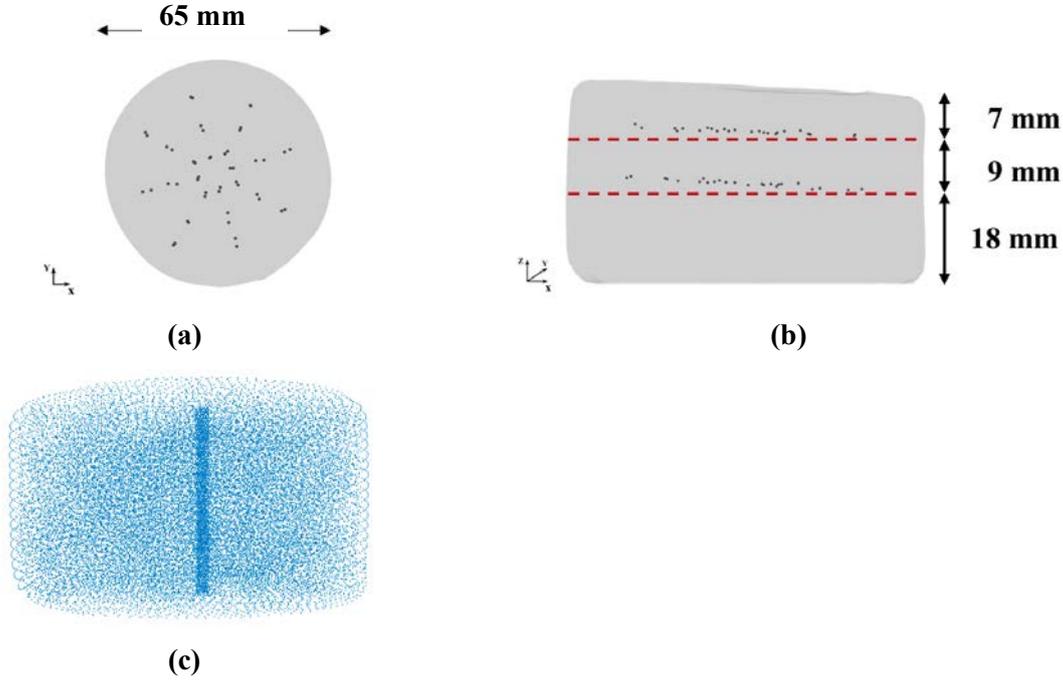

**Fig. 6 (a)** Top and **(b)** side view (X-ray image) of the three-layered non-homogenous (each layer has different material properties) cylindrical Sylgard 527 silicone gel sample with the layers of steel beads (black dots) embedded within the sample; **(c)** computational grid /nodal distribution (32,276 nodes and 178,993 integration cells) we apply to represent the spatial domain when modeling needle insertion into the sample.

**Table 2** Ogden material model parameters for the three layers of the non-homogenous cylindrical Sylgard 527 gel sample. The sample is shown in Fig. 5. The symbols and Ogden model parameters are defined in section *2.2.2 Material model*.

|         | $a$  | $\mu$ (Pa) | $D(Pa^{-1})$         |
| ------- | ---- | ---------- | -------------------- |
| Layer 1 | -1.3 | 1,153      | $3.49249 \times 10^{-5}$ |
| Layer 2 | -1.3 | 1,000      | $4.02684 \times 10^{-5}$ |
| Layer 3 | -1.3 | 866        | $4.64993 \times 10^{-5}$ |

To enable conducting the experiments on needle insertion inside the XCT scanner, we constructed an in-house CT-compatible (manufactured from hard plastic) test apparatus (Fig. 7). The apparatus was not equipped with a load-cell as we observed that the load-cell we applied to measure the force acting on the needle and other load-cells available to us are opaque to the X-ray beam generated by the XCT scanner and induce strong artefacts, making acquisition of high-quality 3D images of the gel sample impossible.

To model needle insertion conducted using the experimental set-up shown in Fig. 6, we use a nodal distribution of 32,276 nodes and 178,993 integration cells (with one integration point per cell) (Fig. 5c), which according to the convergence analysis we conducted in section *3.1.1 Verification of method convergence*, ensures a grid independent (converged) numerical solution.

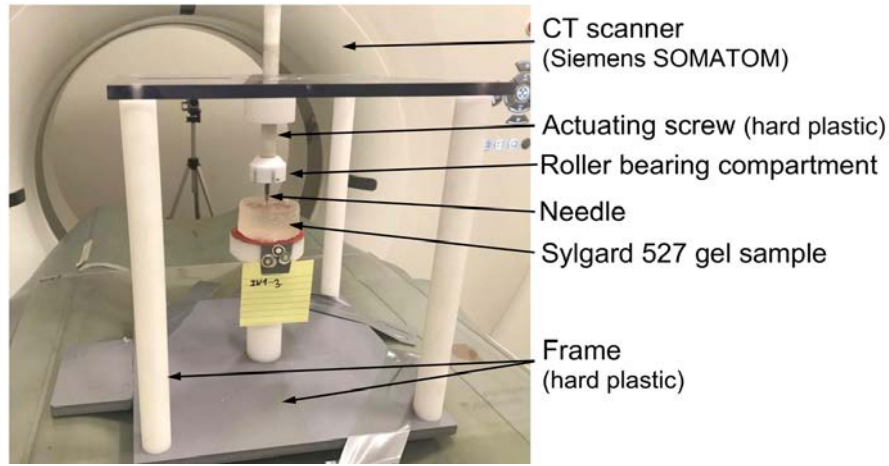

**Fig. 7** Experimental set-up for conducting the experiments on needle insertion within a CT scanner. The experiments were conducted to obtain quantitative information about the deformation field induced by needle insertion. We used an in-house CT-compatible test apparatus built from hard plastic. The needle is attached to a roller bearing located beneath the actuating screw to prevent transmission of the rotary motion of the screw to the needle (i.e. the needle undergoes only translational/linear motion). For CT image acquisition, we used Siemens SOMATOM XCT scanner located at Medical XCT Facility of Commonwealth Scientific and Industrial Research Organization (CSIRO) in Kensington, Western Australia.

**2.5   Needle insertion into continua with complex geometry**

To examine the applicability of the proposed methods for needle insertion modeling to continua with complex geometries, we model needle insertion into a brain phantom geometry determined from the radiographic images (magnetic resonance and computed tomography) as described in (68) (Fig. 8). To discretize the analyzed geometry (domain dimensions of 0.14 m × 0.156 m × 0.14 m) we use 73,926 nodes and 417,790 integration cells (with one integration point per cell) (Fig. 9). All nodes defining the inferior part of the brain phantom outer surface are rigidly constrained (red nodes in Fig. 9). With the exception of the needle insertion point, the remaining outer surface nodes were defined as free (no forces and no displacements prescribed).

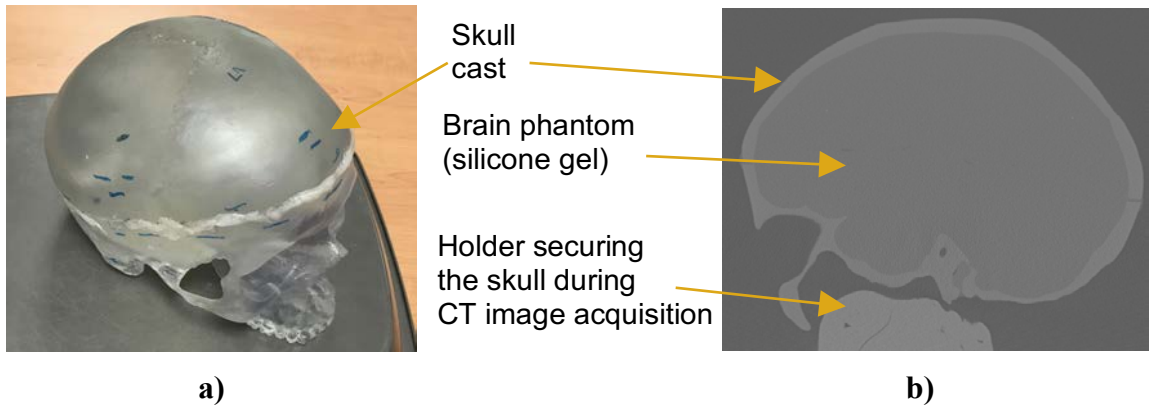

**Fig. 8** Complex geometry (brain phantom made from Sylgard 527 silicone gel) we used in this study to evaluate the performance of our algorithm for needle insertion simulation. **(a)** Photograph of the anatomically accurate human skull cast by 3B Scientific (Hamburg, Germany; https://www.3bscientific.com) we used to manufacture the phantom (inside the skull). **(b)** Computed tomography (CT) image (sagittal section) of the brain phantom. To extract information about the phantom geometry from the images, we used 3D Slicer — an open source software platform for image processing and three-dimensional visualization (69).

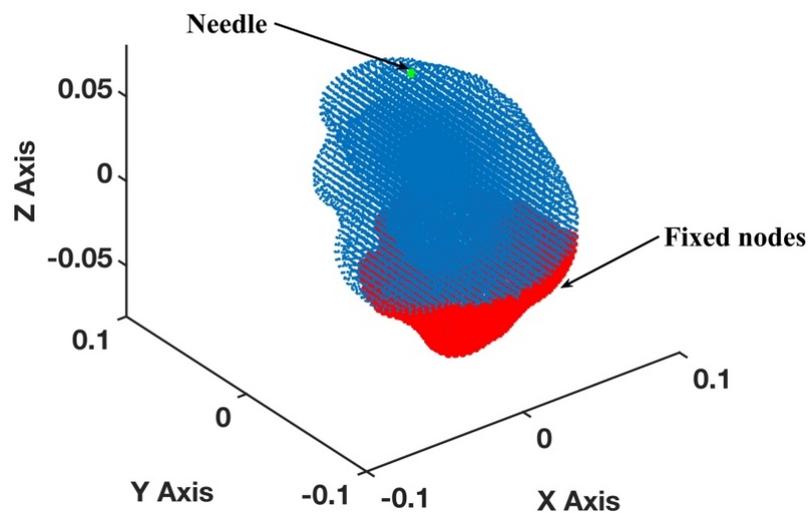

**Fig. 9** Meshless discretization (using the MTLED algorithm) for simulation of the needle insertion into the brain phantom geometry extracted from the phantom radiographic images. The geometry was discretized using 73,926 nodes (blue and red dots) and 417,790 background tetrahedral integration cells with one Gauss point per cell. Red dots indicate the nodes that are rigidly constrained.

## 3. RESULTS

### 3.1 Method verification

We verify our modeling and simulation method by considering needle insertion (diameter of 1.6 mm) into a homogeneous cylindrical sample (diameter 30 mm; height 17 mm; referred to as a *small sample*) made from Sylgard 527 gel (see Fig. 4).

*3.1.1 Verification of method convergence*

The nodal displacements obtained from the coarse (7,480 nodes) and moderate (17,730) grids were interpolated on the refined grid (30,294 nodes) using the interpolating moving least squares (70). The values obtained through such interpolation were applied to investigate the differences between the predicted displacements when increasing the number of nodes. The results are presented in the histograms plots of the-node-by-node differences between the displacement fields obtained using the refined (30,294 nodes) grid, and the displacement fields computed using the coarse (7,480 nodes) (Fig. 9a) and moderate (17,730 nodes) (Fig. 10b) grids interpolated on the refined grid. Practically negligible differences (below 0.1 mm for the vast majority of the grid nodes) between the displacements computed using moderate and refined grids are observed (Fig. 10). This observation is confirmed by the quantitative analysis using the Normalized Root Mean Square Error $NRMSE = \frac{\frac{1}{N}\sqrt{\sum_{i=1}^{N}\left(u_i^{interpolated}-u_i^{denser}\right)^2}}{u_{max}^{denser}-u_{min}^{denser}}$, where $N$ is the number of nodes used the spatial domain discretization, $u_i^{interpolated}$ is the nodal displacement component $(u_x, u_y, u_z)$ is the nodal displacement component obtained by interpolating the results obtained using the coarse and moderate density grids on the dense grid nodes, $u_i^{denser}$ is the nodal displacement component computed using the refined grid (30,294 nodes).

NRMSE for the coarse, moderate and refined computational grids/nodal distributions used here is reported in Table 3. With the maximum NRMSE (for the displacement component in the Z-axis direction) of under $6.5 \times 10^{-3}$, the displacements obtained using the moderate and dense grids are practically indistinguishable. This indicates convergence of the solution for the moderate (17,730 nodes and 96,038 integration points) computational grid and therefore grids of this density were used in further simulations. It may be noted here that even results obtained with a coarse grid (NRMSE of an order to $10^{-2}$) may be acceptably accurate in practice.

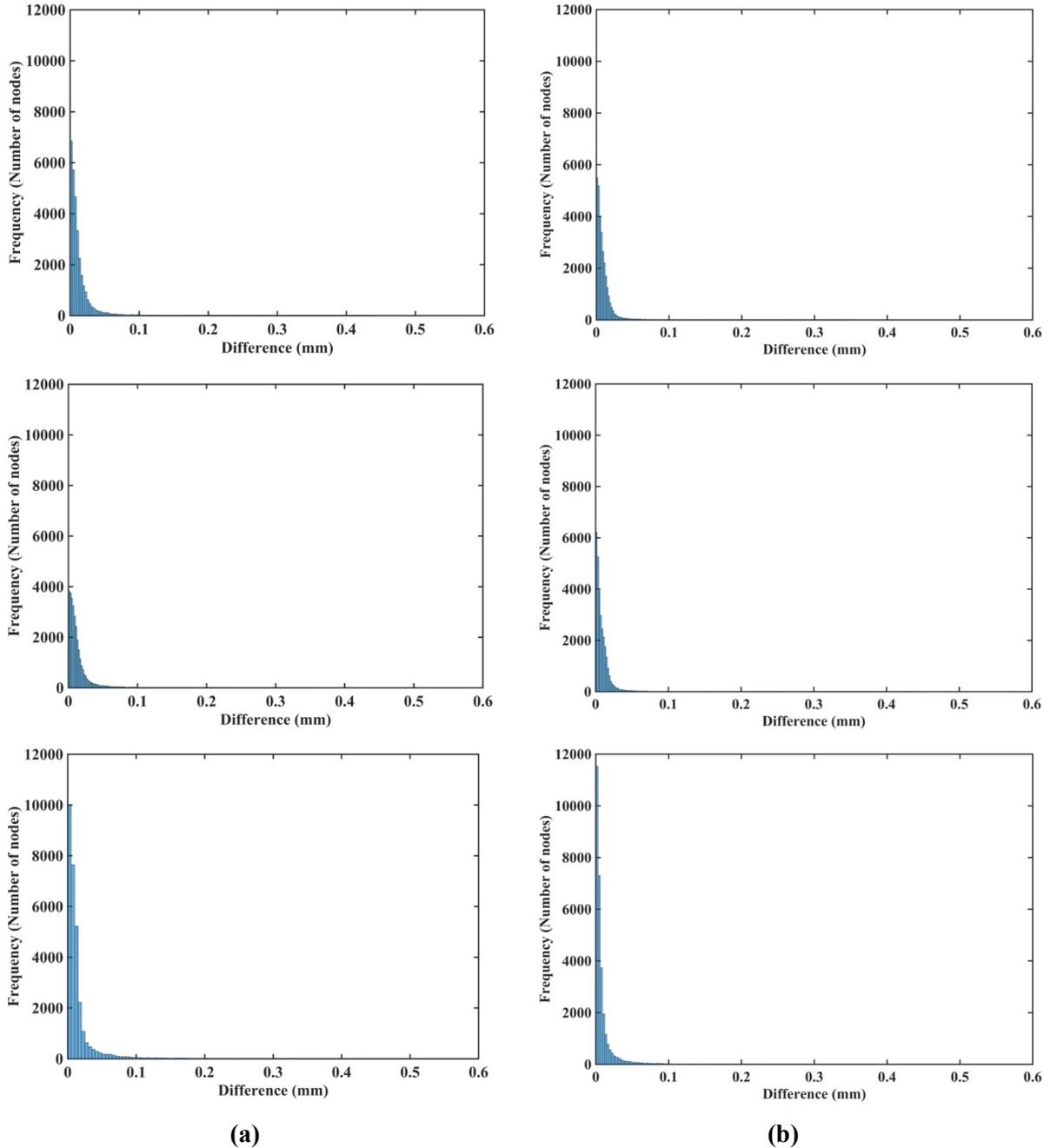

**Fig. 10 (a)** Histograms displaying the differences in displacement field components ($u_x$-top, $u_y$-middle, $u_z$-bottom) between **(a)** the coarse (7,480 nodes) and refined (30,294 nodes) computational grids; and **(b)** the moderate (17,730 nodes) and refined (30,294 nodes) grids. The comparison was done node-by-node for the refined grid. Interpolation was applied to compute the displacements at the locations of nodes of the refined grid using coarse and moderate density grids. The needle is inserted in the *z*-direction. Note practically negligible differences (under 0.1 mm for all the nodes) between the displacement field components obtained using moderate and refined grids.

**Table 3** Normalized root mean square error (NRMSE) for displacement components ($u_x$, $u_y$, $u_z$) for successively denser grids obtained when modeling needle insertion into a cylindrical sample (diameter 30 mm; height 17 mm) of silicone gel. The sample is shown in Fig. 2.

|  | NRMSE | | |
| --- | --- | --- | --- |
|  | $u_x$ | $u_y$ | $u_z$ |
| Coarse (7,480 nodes) to refined (30,294 nodes) grids | $1.36 \times 10^{-2}$ | $1.40 \times 10^{-2}$ | $8.26 \times 10^{-3}$ |
| Moderately dense (17,730 nodes) to refined (30,294 nodes) grids | $9.83 \times 10^{-3}$ | $9.72 \times 10^{-3}$ | $5.75 \times 10^{-3}$ |

*3.1.2 Demonstration of the ability to compute needle reaction force*

The results of simulation of needle insertion into a *small cylindrical gel sample* (diameter 30 mm; height 17 mm; see Fig. 4 for the experimental set-up) confirm that when the mechanical properties of the tissue are known precisely, our method correctly computes not only displacements but also forces and therefore is mechanically consistent (as is nonlinear FEM). The computed force is very close to that experimentally measured (Fig. 11). Slight differences between computed reaction force and experiment can be attributed to the inadequacy of Ogden model at very high strains.

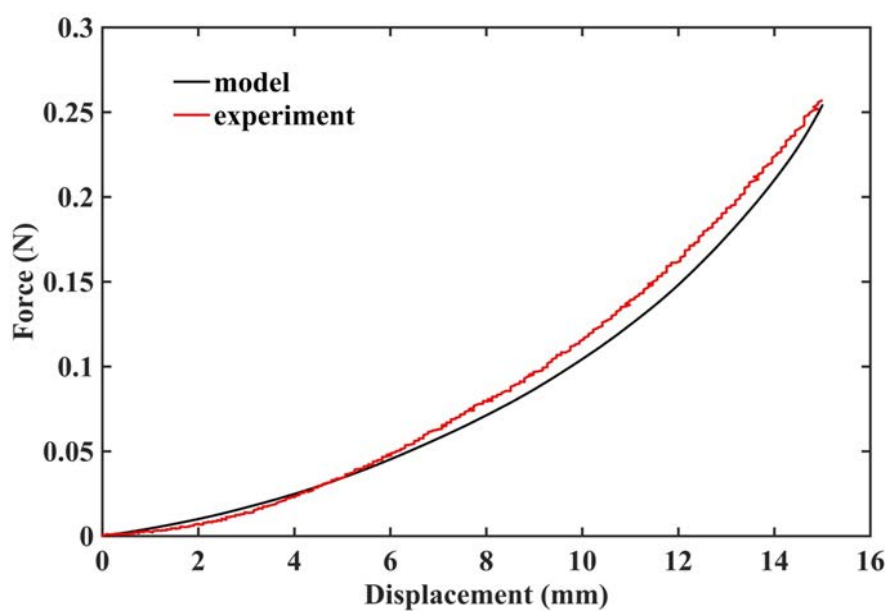

**Fig. 11** Measured (red solid line) and computed (black solid line) force acting on the needle during insertion into *small cylindrical gel sample* (diameter of 30 mm, height of 17 mm). The sample is shown in Fig. 4.

*3.1.3 Demonstration of displacement field independence of material properties*

As indicated in the histograms in Fig. 12, the node-by-node differences between the displacement fields computed when varying the material properties (shear modulus) and material models (Ogden and neo-Hookean) as described in Table 1 are well below 1 μm ($10^{-3}$ mm). This is consistent with the analysis of the Normalized Root Mean Square Error (NRMSE) for the displacement components (see Table 4). As the maximum NMRSE is $4.17 \times 10^{-4}$ (Table 4) for two orders of magnitude shear modulus difference (Material 1 and Material 3), it can be concluded that the displacements computed when varying material models and material properties, as described in Table 1, are for practical purposes indistinguishable. This indicates that the displacement field predicted using our method for needle insertion modeling is independent of the material model and properties used. This feature of our modeling approach is extremely important for patient-specific applications, where we rarely know tissue properties precisely.

**Table 4** Modeling needle insertion into a cylindrical sample (diameter of 30 mm and height of 17 mm) when varying the sample material model and material properties as described in Table 1. The insertion depth is 15 mm. The computational grid consists of 17,730 nodes and 96,038 integration cells with one integration point per cell. Normalized root mean square error (NRMSE) for the displacement components ($u_x$, $u_y$, $u_z$) predicted using different material models and material properties. Materials 1, 2 and 3 are Ogden, and Material 4 is neo-Hookean. Note that two orders of magnitude difference in shear modulus results in negligibly small NRMSE of $4.17 \times 10^{-4}$.

|  | NRMSE | | |
| --- | --- | --- | --- |
|  | $u_x$ | $u_y$ | $u_z$ |
| Material 1 and 2 | $3.85 \times 10^{-4}$ | $4.62 \times 10^{-5}$ | $3.25 \times 10^{-5}$ |
| Material 1 and 3 | $4.17 \times 10^{-4}$ | $5.80 \times 10^{-5}$ | $3.39 \times 10^{-5}$ |
| Material 2 and 4 | $6.33 \times 10^{-5}$ | $1.67 \times 10^{-4}$ | $1.17 \times 10^{-5}$ |

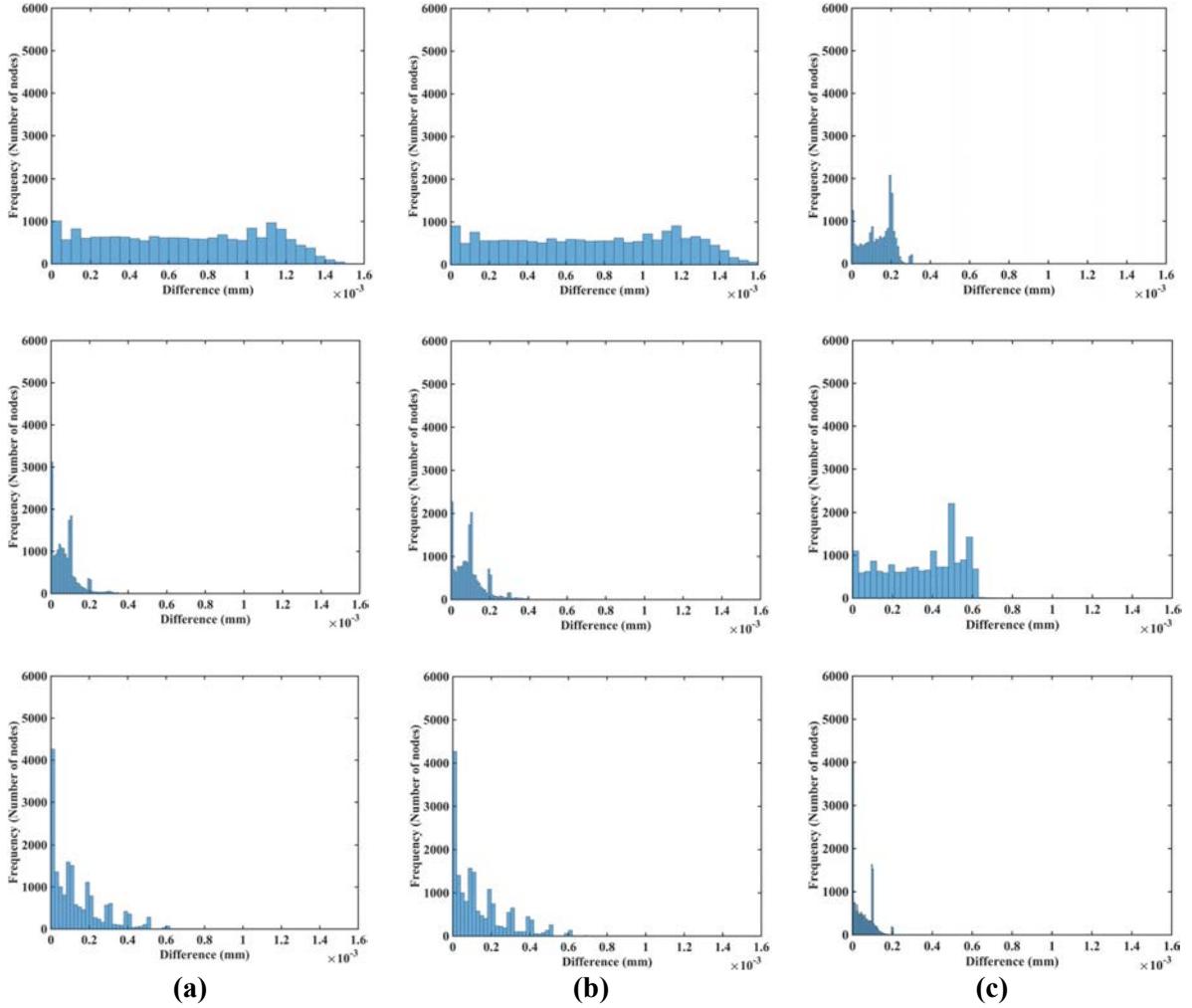

**Fig. 12** Modeling needle insertion into *a small cylindrical sample* (diameter of 30 mm and height of 17 mm; the sample is shown in Fig. 4) when varying the sample material properties (shear modukus) and material model (Ogden and neo-Hookean) as described in Table 1. The insertion depth is 15 mm. The computational grid consists of 17,730 nodes and 96,038 integration cells with one integration point per cell. Histograms show the node-by-node differences (in millimeters) in displacement field components ($u_x$-top, $u_y$-middle, $u_z$-bottom) between **(a)** Materials 1 and 2 and **(b)** between Materials 1 and 3 and **(c)** between Materials 2 and 4. Materials 1, 2 and 3 are Ogden, and Material 4 is neo-Hookean. The shear moduli are 72.0 Pa for Material 1, 722.0 Pa for Material 2, 7222.0 Pa for Material 3, and 1000.0 Pa for Material 4. See Table 1 for more information about Material 1, Material 2, Material 3 and Material 4.

### 3.2 Experimental evaluation of the method

We model the needle insertion to a depth of up to 15 mm into a nonhomogenous cylindrical sample (diameter of 65 mm and height 34 mm) made from silicone gel (Fig. 6). We compute the displacement field and reaction force on the needle shaft. We use a nodal distribution of 32,276 nodes and 178,993 integration cells (with one integration point per cell) (Fig. 6c). To qualitatively evaluate the accuracy of our kinematic approach for needle insertion modeling, we compare the general deformation/shape of the gel sample predicted using our model with the CT images acquired in the experiments conducted under the set-up shown in Fig. 7. For the quantitative evaluation, we compare the displacement field within the sample at the location of the beads predicted using our model with the beads displacements determined from the analysis of the CT images acquired during needle insertion into the gel sample as shown in Fig. 7. The comparison was done for the needle insertion depth of 5 mm and 15 mm.

*3.2.1 Accuracy of prediction of the displacement field during needle insertion: Qualitative evaluation*

The sample general deformation and shape predicted using our model is very close to the CT images acquired during the needle insertion (see Fig. 13 and Fig. 14). This qualitative observation is consistent with the quantitative analysis of the predicted and experimentally determined displacement field within the sample reported in section *3.2.2 Accuracy of prediction of the displacement field during needle insertion: Quantitative evaluation*.

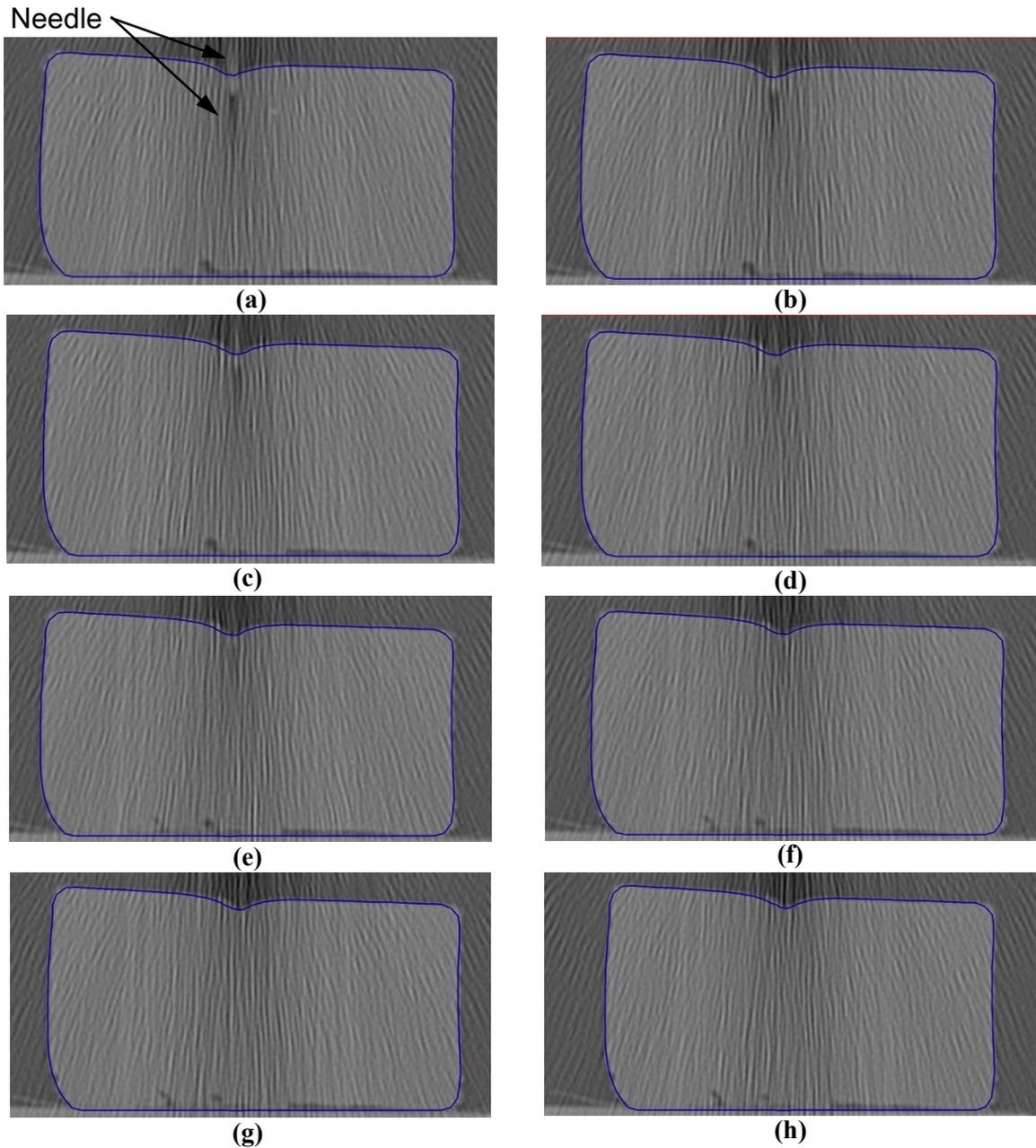

**Fig. 13** Needle insertion (to depth of 5 mm) into the non-homogenous cylindrical sample (diameter of 65 mm and height of 34 mm) of Sylgard 527 gel (Fig. 6 and Fig. 7). The needle diameter is 1.6 mm. The predicted contours of the sample (blue lines) are overlaid on the CT images acquired during the needle insertion. The needle is indicated in figure **(a)**, and its outline can be distinguished in figures **(a)-(d)**. **(a)** Sections through the planes located **(a)** 0 mm. **(b)** 0.16 mm **(c)** 0.32 mm **(d)** 0.48 mm **(e)** 0.64 mm **(f)** 0.80 mm **(g)** 0.96 and **(h)** 1.12 mm anteriorly from the central plane of the needle shaft.

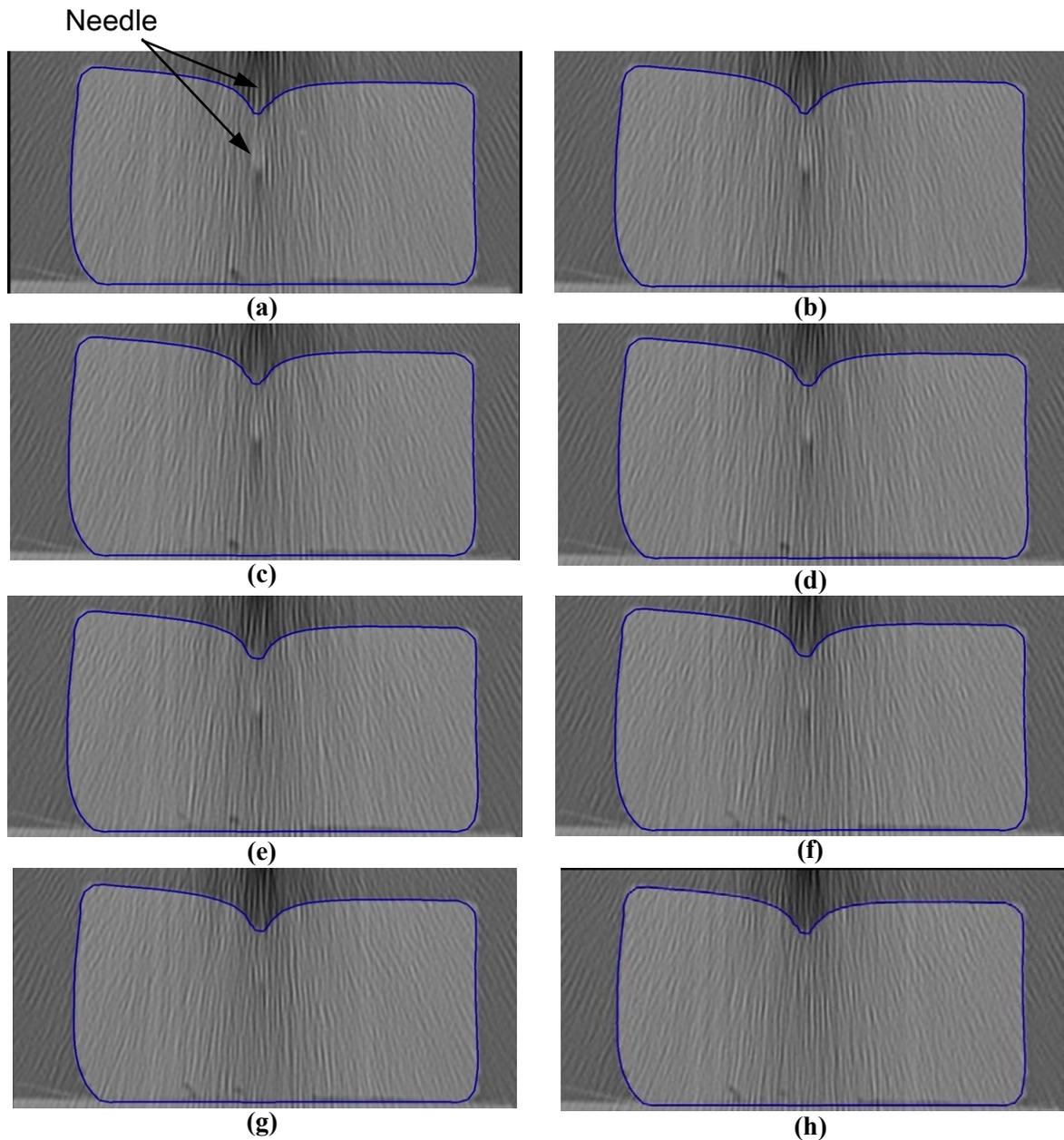

**Fig. 14** Needle insertion (to a depth of 15 mm) into the non-homogenous cylindrical sample (diameter of 65 mm and height of 34 mm) of Sylgard 527 gel (Fig. 6 and Fig. 7). The needle diameter is 1.6 mm. The predicted contours of the sample (blue lines) are overlaid on the CT images acquired during the needle insertion. The needle is indicated in figure **(a)**, and its outline can be distinguished in figures **(a)-(f)**. **(a)** Sections through the planes located **(a)** 0 mm. **(b)** 0.16 mm **(c)** 0.32 mm **(d)** 0.48 mm **(e)** 0.64 mm **(f)** 0.80 mm **(g)** 0.96 and **(h)** 1.12 mm anteriorly from the central plane of the needle shaft.

*3.2.2 Accuracy of prediction of the displacement field during needle insertion: Quantitative evaluation*

As it is difficult to avoid an error smaller than 2 pixels when determining the location of a steel bead in the image, following our previous studies on non-rigid image registration using computational biomechanics models (71-73), we used twice the in-plane image resolution (twice the in-plane pixel size) as the criterion for successful displacement prediction. This implies that when the difference between the predicted and experimentally determined bead displacements does not exceed twice the in-plane voxel size, we regard the prediction as accurate. As the resolution (voxel size) of our images was of 0.16 mm $\times$ 0.16 mm $\times$ 0.10 mm, we use the accuracy criterion of 2 $\times$ 0.16 mm = 0.32 mm for the displacement field components in the *x*-direction and *y*-direction, and 2 $\times$ 0.10 mm = 0.20 mm for the displacement component in the *z*-direction.

Fig. 15 shows the histograms of the differences between the displacement field components (along the three axes of the coordinate system), at the locations of the metallic beads embedded within the sample, predicted using our kinematic approach and measured from the CT images for the needle insertion depth of 5 mm. For 87% (40 out of 46) of the beads the displacement in the *x*-direction was accurately predicted by our model, as the difference between the modeling and experimentally determined displacements does not exceed the 0.32 mm (twice the image resolution) accuracy threshold (Fig. 15). For the *y*-direction and *z*-direction displacement field components the accurate prediction (difference of up to 0.32 mm in the *y*-direction and up to 0.2 mm in the *z*-direction) was achieved for 67% (31 out of 46) beads.

The beads for which the difference between the predicted and experimentally determined displacement magnitude exceeded twice the image resolution (0.32 mm) were located distantly from the needle (Fig. 16). In these areas the image contrast is poor, which introduces errors when determining location of the beads.

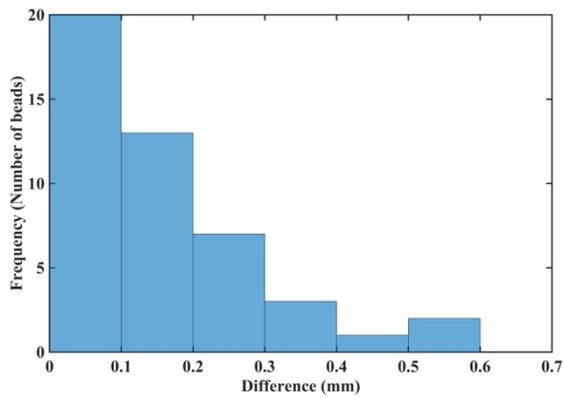
(a)

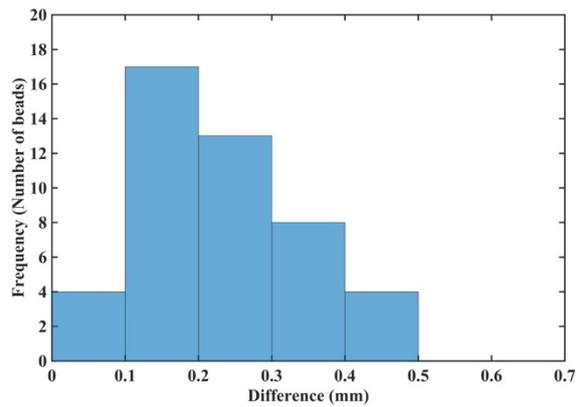
(b)

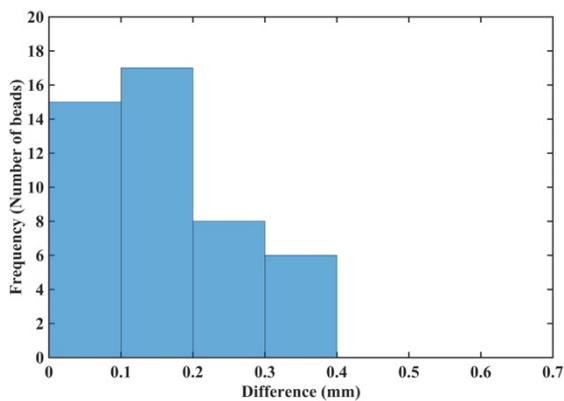
(c)

**Fig. 15** Needle insertion to the depth of 5 mm into the non-homogenous cylindrical sample (diameter of 65 mm and height of 34 mm) of Sylgard 527 gel (Fig. 6 and Fig. 7). Comparison of the displacement field components at the beads location predicted using the MTLED algorithm with the kinematic approach for needle insertion modeling we introduced in this study and experimentally determined from the CT images. Histograms of the differences in the **(a)** $x$-direction, **(b)** $y$-direction and **(c)** z-direction. The displacements are in millimeters. The predicted and experimentally determined displacement field magnitudes for all beads are listed in Table A1 in the Appendix.

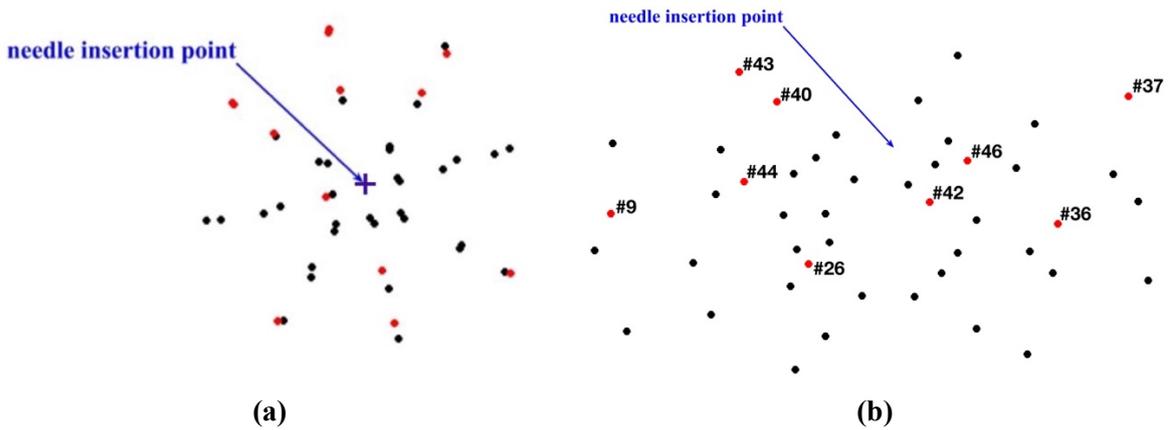

**Fig. 16** Needle insertion to the depth of 5 mm into the non-homogenous cylindrical sample (diameter of 65 mm and height of 34 mm) of Sylgard 527 gel (Fig. 6 and Fig. 7). **(a)** Top view (X-Y plane) and **(b)** Transverse view (X-Z plane) of the steel beads (black and red solid circles) embedded in the gel sample as shown in Fig. 5a and Fig. 5b. Black circles: the difference between the bead displacements predicted using the MTLED algorithm with the kinematic approach for needle insertion modeling and experimentally determined from the CT images does not exceed 0.32 mm (twice the in-plane-image resolution). Red circles with numbers (bead ID): the difference between the bead displacements predicted using the MTLED algorithm with the kinematic approach for needle insertion modeling and experimentally determined from the CT images exceeds 0.32 mm (twice the in-plane-image resolution).

The results of quantitative analysis of the accuracy of prediction of the gel sample deformations for the needle insertion depth of 15 mm are consistent with those for the insertion depth of 5 mm. For the insertion depth of 15 mm, the displacement in the *x*-direction was accurately predicted for 72% (33 out of 46) beads (Fig. 17a), and the displacement in the *y*-direction — for 87% (40 out of 46) beads (Fig. 14b). Although for the displacement in the *z*-direction, the accurate prediction was achieved only for 44% (20 beads out of 46) (Fig. 17c).

As with modeling needle insertion to the depth of 5 mm, majority of the beads for which the difference between the predicted and experimentally determined displacement magnitude exceeded 0.32 mm was located distantly from the needle (Fig. 18), where the image contrast is poor. This tendency is also clearly visible in the vector plot of the predicted and image-determined displacements of the beads (Fig. 19).

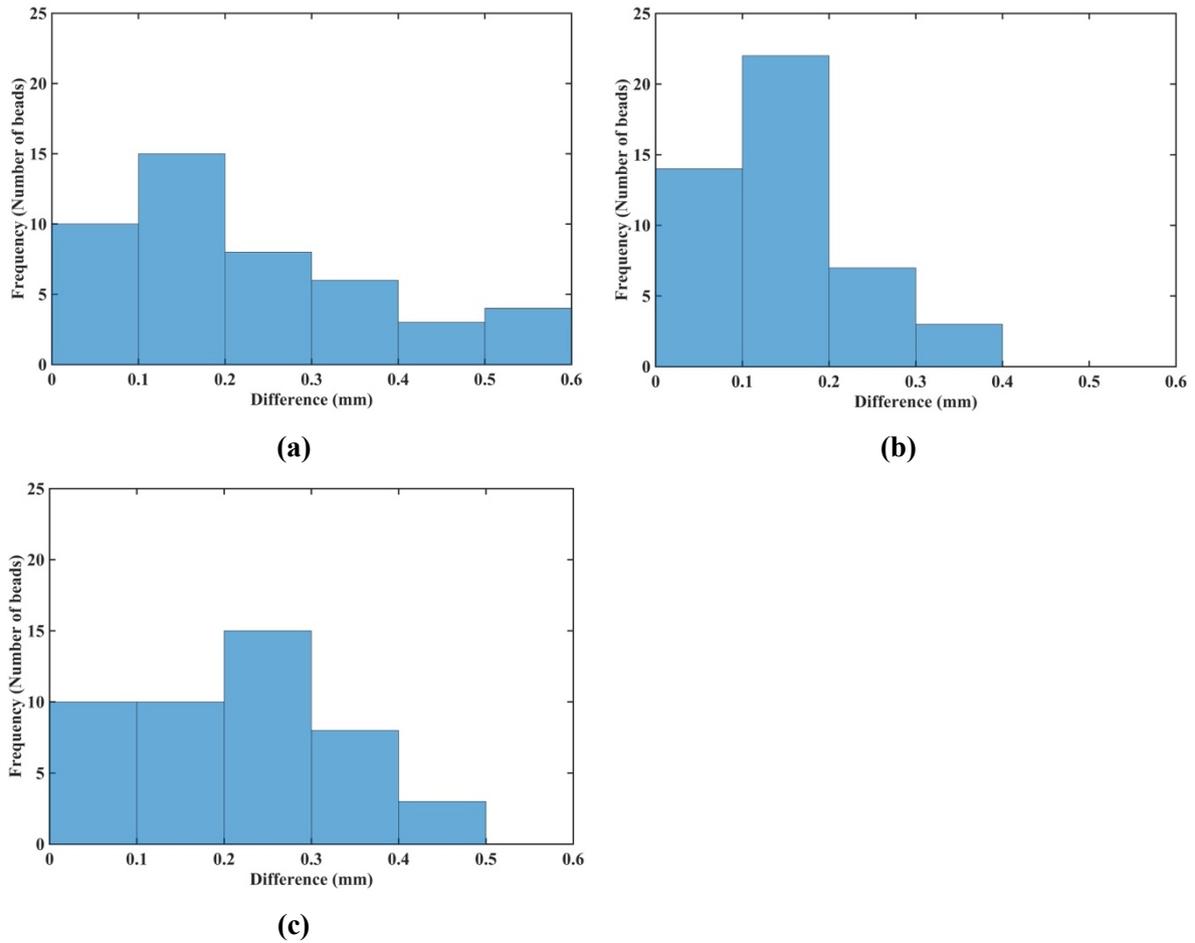

**Fig. 17** Needle insertion to the depth of 15 mm into the non-homogenous cylindrical sample (diameter of 65 mm and height of 34 mm) of Sylgard 527 gel (Fig. 6 and Fig. 7). Comparison of the displacement field components at the beads location predicted using the MTLED algorithm with the kinematic approach for needle insertion modeling we introduced in this study and experimentally determined from the CT images. Histograms of the differences in the **(a)** *x*-direction, **(b)** *y*-direction, and **(c)** z-direction. The displacements are in millimeters. Values of the predicted and experimentally determined displacement field magnitude for all beads are listed in Table A2 in the Appendix.

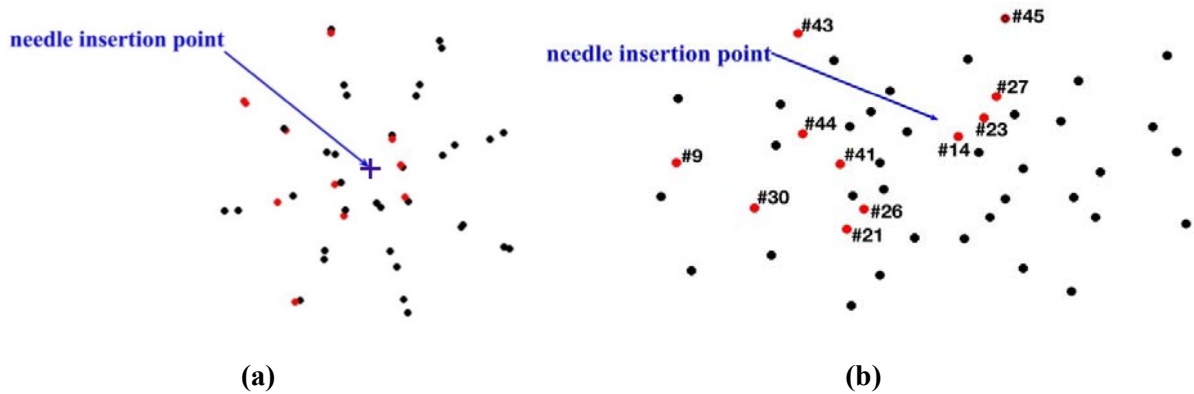

**Fig. 18** Needle insertion to the depth of 15 mm into the non-homogenous cylindrical sample (diameter of 65 mm and height of 34 mm) of Sylgard 527 gel (Fig. 6 and Fig. 7). **(a)** Top view (X-Y plane) and **(b)** Transverse view (X-Z plane) of the steel beads (black and red solid circles) embedded in the gel sample as shown in Fig. 5a and Fig. 5b. Black circles: the difference between the bead displacements predicted using the MTLED algorithm with the kinematic approach for needle insertion modeling and experimentally determined from the CT images does not exceed 0.32 mm (twice the in-plane-image resolution). Red circles with numbers (bead ID): the difference between the bead displacements predicted using the MTLED algorithm with the kinematic approach for needle insertion modeling and experimentally determined from the CT images exceeds 0.32 mm (twice the in-plane-image resolution).

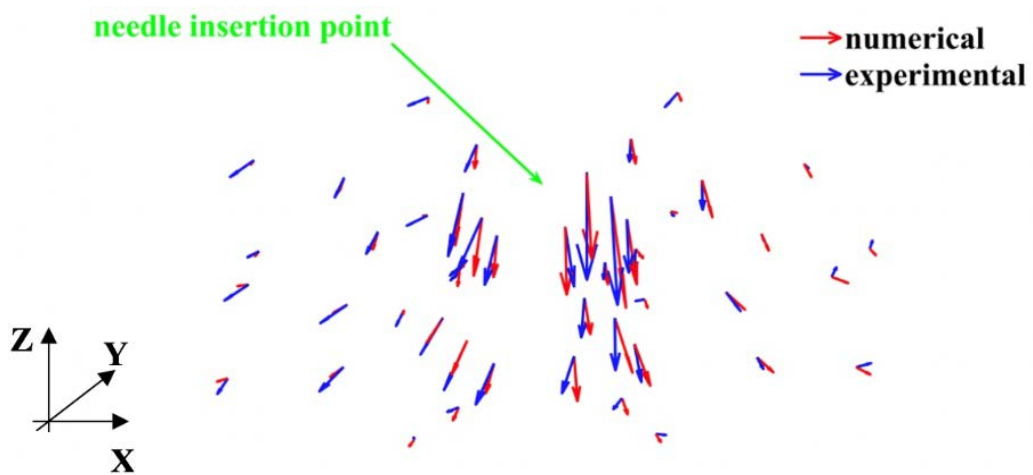

**Fig. 19** Modeling of needle insertion (insertion depth of 15 mm) into the non-homogenous cylindrical (diameter of 65 mm and height of 34 mm) sample of Sylgard 527 gel. Vector plot of the predicted (red vectors) using our model and experimentally determined from the CT image analysis (blue vectors) displacement field at the beads' location. The gel sample is shown in Fig. 6, and the experimental set-up — in Fig. 7.

*3.2.3 Prediction of force acting on the needle*

The predicted and experimentally measured (the experimental set-up is shown in Fig. 4) forces acting on the needle during insertion into the non-homogenous cylindrical sample (diameter of 65 mm and height of 34 mm) of Sylgard 527 gel agree very well for the insertion depth of up to 10 mm (Fig. 20). However, the differences between the modeling and experimental results increase with the needle insertion depth. One possible explanation for this tendency can be that the Ogden model obtained from the experiments (see Fig. 1a) loses its accuracy at very high strains.

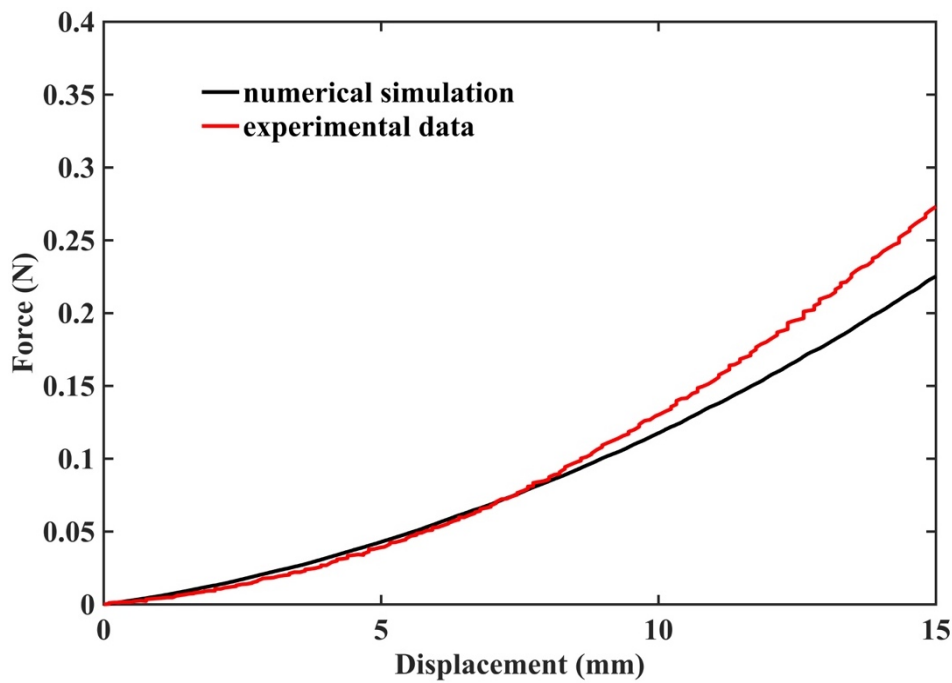

**Fig. 20** Measured (red solid line) and predicted (black solid dashed) force acting on the needle during insertion into the non-homogenous cylindrical sample (diameter of 65 mm and height of 34 mm) of Sylgard 527 gel. The sample and the model are shown in Fig. 6. The experimental set-up is shown in Fig. 4.

*3.2.4 Weak dependence on mechanical properties*

To demonstrate that our approach is effective in predicting tissue deformations even when its mechanical properties are unknown, we repeated the needle insertion simulation into a large sample (diameter of 65 and height of 34 mm) with beads using uniformly the simplest neo-Hookean model with $\mu = 1,000$ Pa.

Fig. 21, shows the histogram plot of the node-by-node differences between the displacement fields computed using the Ogden material model (three-layers with material properties listed in Table 2) and neo-Hookean material model with the uniform properties for the entire sample. The Normalized Root Mean Square Error (NRMSE) for the $u_x, u_y$ and $u_z$ displacement components is $3.1 \times 10^{-3}$, $8.1 \times 10^{-3}$, and $3.2 \times 10^{-3}$, respectively. This result confirms that for image-guided surgical operations, where prediction of displacements is of importance, our method can be used without the knowledge of patient-specific properties of tissues.

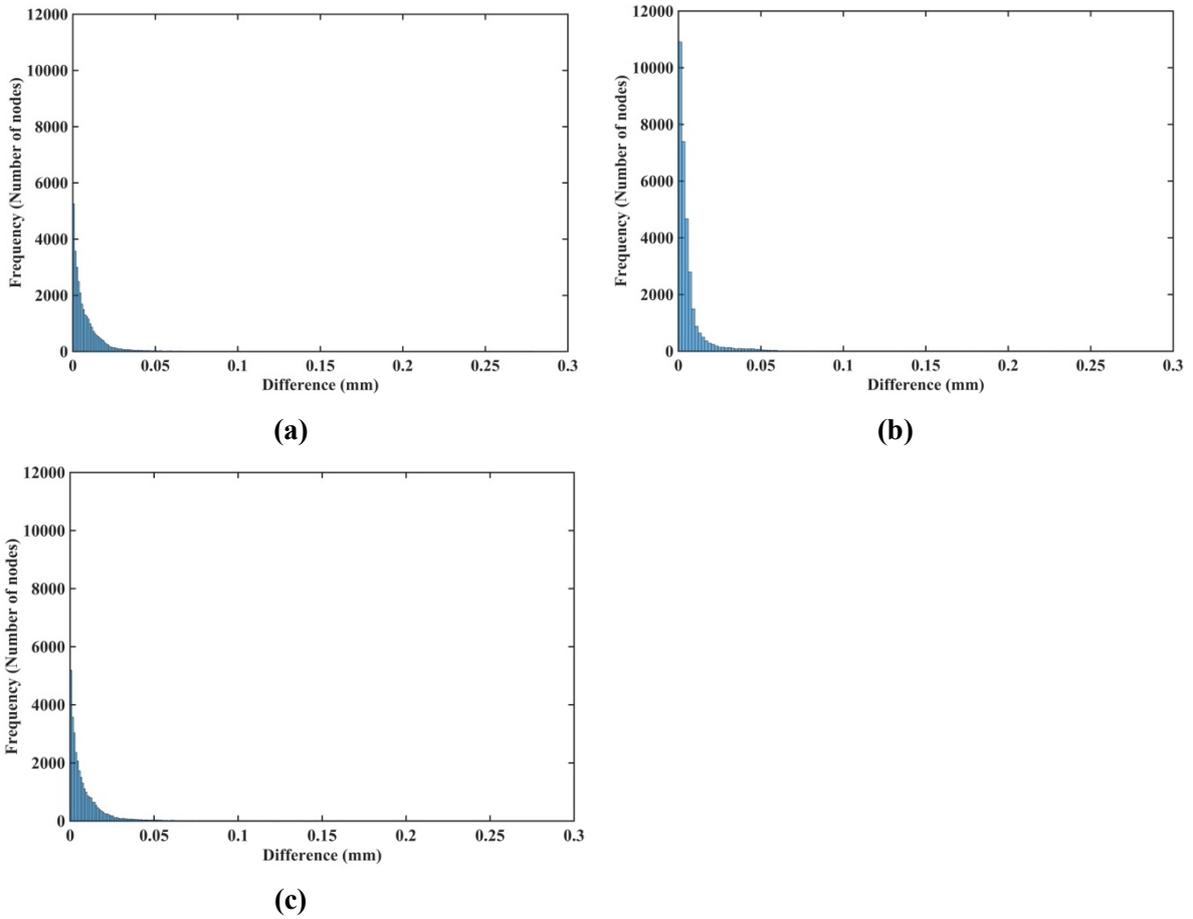

**Fig. 21** Modeling needle insertion into a cylindrical sample (diameter of 65 mm and height of 34 mm) using uniformly the simplest neo-Hookean model with $\mu$ = 1,000 Pa. The insertion depth is 15 mm. Histograms displaying the node-by-node difference (in mm) for the **(a)** $u_x$ **(b)** $u_y$ and **(c)** $u_z$ displacement field components between Ogden and neo-Hookean material models. In the simulations using the Ogden material model, three layers with different material properties were distinguished in the sample as listed in Table 2. For the neo-Hookean material model, the sample was modeled as a homogenous continuum (uniform material properties for the entire sample). Note practically negligible differences (up to 0.05 mm for all the nodes) between the displacements obtained using the two material models.

### 3.3 Needle insertion into continua with complex geometry

We model the needle insertion into the brain phantom geometry shown in Fig. 8 and Fig. 9. The insertion was conducted to a depth of 100 mm (in z-direction in Fig. 22). The parameters for the Ogden material model used are the same as those used section *3.1 Method verification* when modeling needle insertion into *the small cylindrical sample* (diameter of 30 mm and height of 17 mm) of silicone gel (see Fig. 4). The needle diameter (1.6 mm) and the deformation coefficient ($C_D$=0.4) are the same as in the simulations conducted in section *3.1 Method verification* and section *3.1 Experimental verification of the method*.

Magnitude of the displacement field at the insertion depth of 100 mm is shown in Fig. 22. The simulation was conducted without any instabilities encountered.

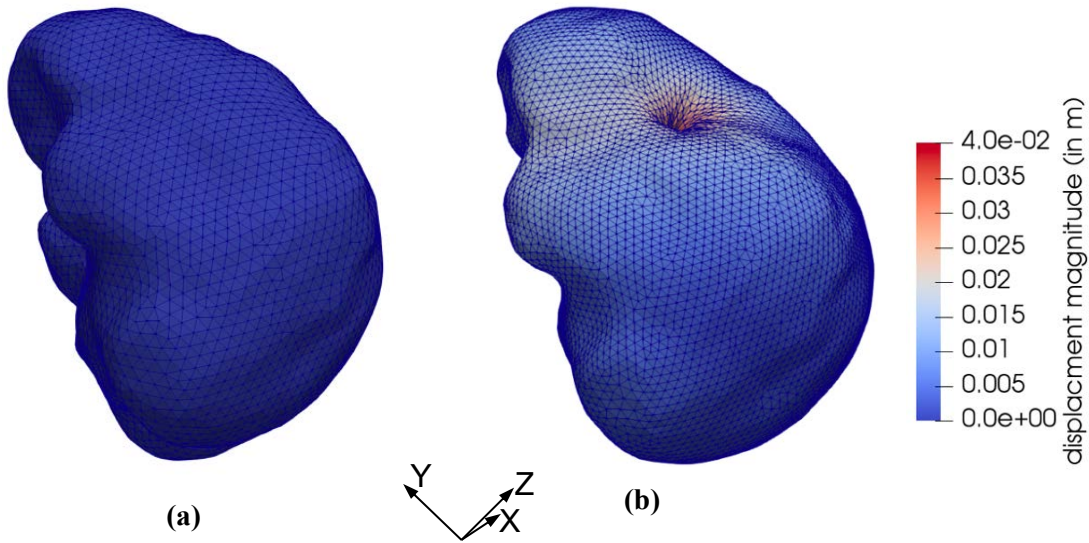

**Fig. 22 (a)** Initial configuration of the brain phantom geometry (the geometry is shown in Fig. 8 and Fig. 9) and **(b)** Magnitude of the displacement field (in m) when modeling needle insertion by 0.10 m in *z*-direction.

To demonstrate that our approach is effective in predicting tissue deformations even when its mechanical properties are unknown, we repeated our simulation using the simplest neo-Hookean model with $\mu$ = 1,000 Pa. Fig. 23, shows the histogram plot of the node-by-node differences between the displacement fields computed using the realistic Ogden material model and the simplest neo-Hookean model. The Normalized Root Mean Square Error (NRMSE) for the $u_x, u_y$ and $u_z$ displacement components is $3.2 \times 10^{-2}$, $3.5 \times 10^{-5}$, and $1.91 \times 10^{-2}$, respectively.

These results should be interpreted in the context of image-guided surgery that often rely on intraoperative Magnetic Resonance Images (MRIs). Resolution of intraoperative MRIs is typically not better than 1 mm (73) and the desired accuracy of many surgical procedures is of an order of 1-2 mm (below 2 mm for the brain tumor resection) (74). Therefore, the results shown in Fig. 23 can be interpreted as confirmation that the results of prediction of the displacement field due to needle insertion obtained using our methods are, for practical purposes, independent of the knowledge/assumptions about the mechanical properties of the analyzed tissue (continuum). Our methods may facilitate sufficiently accurate prediction of the displacement field even without exact knowledge of such properties.

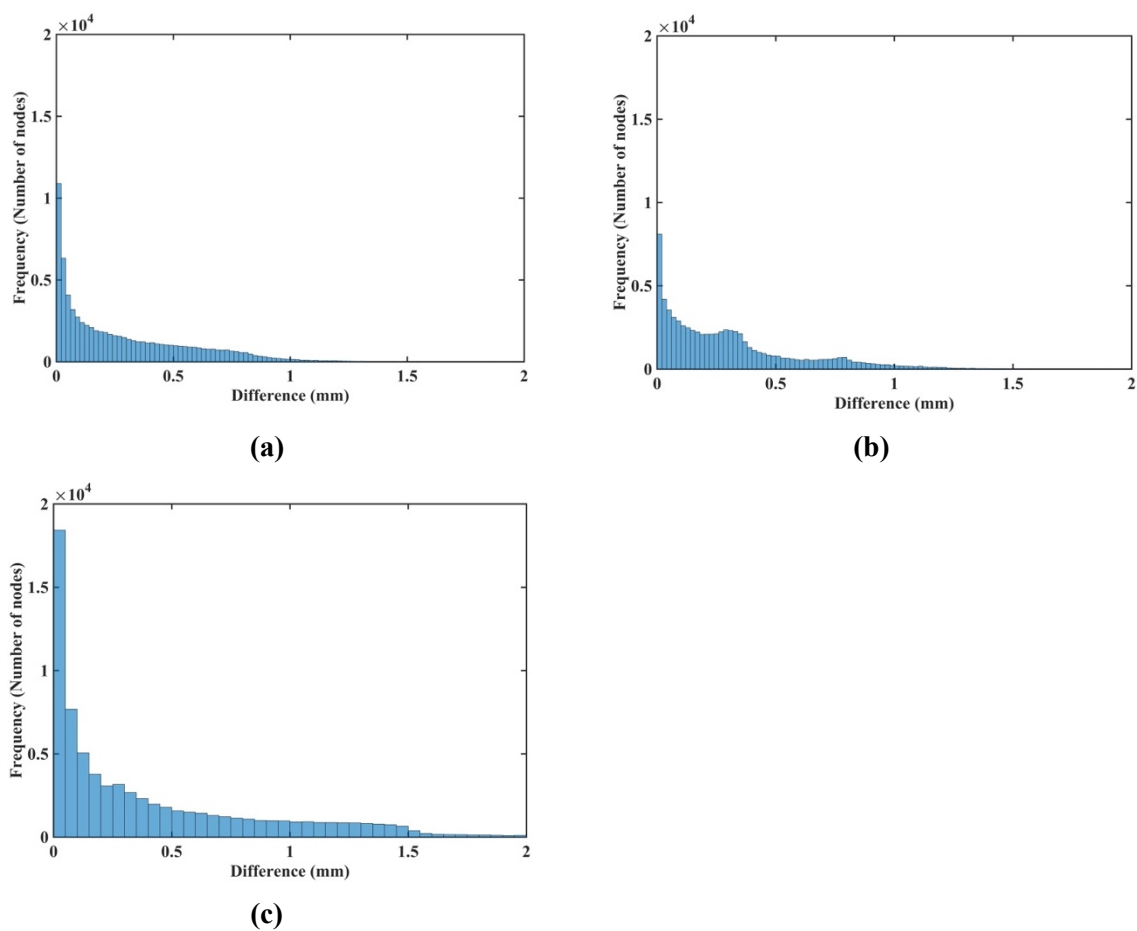

**Fig. 23** Modeling of needle insertion into a continuum with complex geometry (the brain phantom – see Fig. 8) using Ogden (with $\mu$ =722 Pa and $\alpha$ = -1.3) and the simplest hyperelastic neo-Hookean material models with $\mu$ =1,000 Pa. The insertion depth is 100 mm. Histograms displaying the node-by-node difference (in mm) for the **(a)** $u_x$ **(b)** $u_y$ and **(c)** $u_z$ displacement field components between Ogden and neo-Hookean material models. The brain phantom geometry was discretized using 73,926 nodes and 417,790 background tetrahedral integration cells with one integration point per cell as shown in Fig. 9.

# 4. Discussion

In the present contribution we simulate needle insertion into soft tissues. We use a fully geometrically and materially non-linear solid mechanics framework. To avoid the requirement for patient-specific material properties of the tissue, as well as tissue-needle interaction models, we propose a novel kinematics-based modeling approach. Our kinematic approach is consistent with our long-held belief that only methods that do not depend on the knowledge of patient-specific material parameters have a prospect of being successfully translated to the clinic (48-50, 66). Our modeling method requires only two parameters that are easy to identify from images.

To compute soft tissue and other soft continua deformations our method uses the Meshless Total Lagrangian Explicit Dynamics (MTLED) suite of algorithms (31). Our meshless method has very significant advantages as compared to commonly used finite element method. Labor intensive and time consuming finite element meshing is totally eliminated, making our meshless method potentially compatible with clinical workflows (32). Moreover, our meshless approach can handle very large deformations and strains (common near the contact of a surgical tool and a soft organ) effortlessly (31), while the finite element method fails in such cases unless expensive re-meshing is employed (32).

We envisage the primary applications of our new methodology in navigation for image-guided surgery and surgical simulation. The primary variable of interest there is the displacement field within a soft organ. In this paper we demonstrated that our methods are suitable for accurate computation of a displacement field caused by needle motion without patient-specific material models of tissues and without detailed modeling of needle-tissue interaction. Moreover, we demonstrated that when the mechanical properties of modeled continuum are known, our methods accurately recover also reaction force on the needle.

Our needle insertion simulation results would be difficult to replicate with any other numerical method.

**Acknowledgements**

This research was supported by the Australian Government through the Australian Research Council's *Discovery Projects* funding scheme (project DP160100714).

We thank Mr Clement Blondin, a visiting research student (from CAM Graduate Engineering School in Lille, France) at Intelligent Systems for Medicine Laboratory at the University of Western Australia, for help in analyzing the CT images of Sylgard 527 gel samples undergoing needle insertion.

We thank Mr Augustine Cloix, a visiting research student (from SIGMA Engineering Graduate School in Clermont-Ferrand, France) at Intelligent Systems for Medicine Laboratory at the University of Western Australia, for help in analysis of the forces obtained when subjecting Sylgard 527 gel samples to semi-confined compression.

We thank Mr Johann D. Visser, Master of Professional of Engineering student at the Intelligent Systems for Medicine Laboratory at the University of Western of Australia, for preparing Sylgard 527 gel samples used in this study.

# APPENDIX

**Table A1** Predicted (using the MTLED algorithm with the kinematic approach for needle insertion modeling we introduced in this study) and experimentally determined (from the CT images) displacement field magnitude of the beads for the needle insertion to the depth of 5 mm. Numbers in red indicate the difference between the predicted and experimentally determined displacement magnitude greater than 0.32 mm (twice the in-plane image resolution). This table lists the numerical values for the results shown in Fig. 15 and Fig. 16.

| # of beads | Numerical (mm) | Experimental (mm) | \|Difference\| (mm) |
|---|---|---|---|
| 1 | 0.088 | 0.078 | 0.010 |
| 2 | 0.100 | 0.142 | 0.041 |
| 3 | 0.117 | 0.143 | 0.025 |
| 4 | 0.085 | 0.069 | 0.016 |
| 5 | 0.150 | 0.310 | 0.159 |
| 6 | 0.153 | 0.263 | 0.109 |
| 7 | 0.135 | 0.306 | 0.171 |
| 8 | 0.173 | 0.177 | 0.321 |
| 9 | 0.056 | 0.440 | 0.383 |
| 10 | 0.084 | 0.169 | 0.084 |
| 11 | 0.115 | 0.385 | 0.270 |
| 12 | 0.122 | 0.407 | 0.284 |
| 13 | 0.244 | 0.182 | 0.061 |
| 14 | 0.495 | 0.376 | 0.118 |
| 15 | 0.141 | 0.279 | 0.137 |
| 16 | 0.140 | 0.415 | 0.275 |
| 17 | 0.233 | 0.448 | 0.214 |
| 18 | 0.499 | 0.274 | 0.225 |
| 19 | 0.318 | 0.333 | 0.014 |
| 20 | 0.308 | 0.233 | 0.075 |
| 21 | 0.190 | 0.485 | 0.295 |
| 22 | 0.196 | 0.103 | 0.093 |
| 23 | 0.779 | 0.637 | 0.140 |
| 24 | 0.287 | 0.376 | 0.088 |
| 25 | 0.438 | 0.490 | 0.051 |
| 26 | 0.210 | 0.543 | 0.333 |
| 27 | 0.686 | 0.754 | 0.069 |
| 28 | 0.146 | 0.272 | 0.126 |
| 29 | 0.255 | 0.326 | 0.071 |
| 30 | 0.110 | 0.324 | 0.213 |
| 31 | 0.416 | 0.448 | 0.030 |
| 32 | 0.062 | 0.292 | 0.229 |
| 33 | 0.231 | 0.421 | 0.189 |
| 34 | 0.195 | 0.473 | 0.278 |
| 35 | 0.079 | 0.289 | 0.210 |
| 36 | 0.165 | 0.547 | 0.382 |
| 37 | 0.102 | 0.688 | 0.586 |
| 38 | 0.215 | 0.378 | 0.163 |
| 39 | 0.119 | 0.202 | 0.082 |
| 40 | 0.123 | 0.549 | 0.426 |
| 41 | 0.125 | 0.435 | 0.310 |
| 42 | 0.151 | 0.572 | 0.420 |
| 43 | 0.066 | 0.682 | 0.616 |
| 44 | 0.089 | 0.507 | 0.418 |
| 45 | 0.081 | 0.145 | 0.064 |
| 46 | 0.103 | 0.560 | 0.457 |

**Table A2** Predicted (using the MTLED algorithm with the kinematic approach for needle insertion modeling we introduced in this study) and experimentally obtained (from the CT images) displacement field magnitude of the beads for the needle insertion to the depth of 15 mm. Numbers in red indicate the difference between the predicted and experimentally determined displacement magnitude greater than 0.32 mm (twice the in-plane image resolution). This table lists the numerical values for the results shown in Fig. 17 and Fig. 18.

| # of beads | Numerical (mm) | Experimental (mm) | \|Difference\| (mm) |
|---|---|---|---|
| 1 | 0.243 | 0.248 | 0.005 |
| 2 | 0.404 | 0.320 | 0.084 |
| 3 | 0.457 | 0.405 | 0.052 |
| 4 | 0.253 | 0.246 | 0.007 |
| 5 | 0.515 | 0.215 | 0.300 |
| 6 | 0.623 | 0.384 | 0.238 |
| 7 | 0.563 | 0.602 | 0.039 |
| 8 | 0.618 | 0.443 | 0.174 |
| 9 | 0.145 | 0.485 | <span style="color:red">0.340</span> |
| 10 | 0.348 | 0.300 | 0.048 |
| 11 | 0.487 | 0.564 | 0.076 |
| 12 | 0.423 | 0.341 | 0.081 |
| 13 | 1.242 | 1.020 | 0.221 |
| 14 | 1.885 | 1.430 | <span style="color:red">0.455</span> |
| 15 | 0.469 | 0.317 | 0.151 |
| 16 | 0.544 | 0.391 | 0.153 |
| 17 | 1.254 | 1.072 | 0.182 |
| 18 | 1.921 | 1.613 | 0.308 |
| 19 | 1.163 | 0.879 | 0.283 |
| 20 | 1.179 | 1.336 | 0.157 |
| 21 | 0.890 | 1.216 | <span style="color:red">0.325</span> |
| 22 | 0.827 | 0.713 | 0.113 |
| 23 | 3.326 | 2.947 | <span style="color:red">0.378</span> |
| 24 | 1.729 | 1.420 | 0.308 |
| 25 | 1.692 | 1.760 | 0.067 |
| 26 | 1.051 | 1.478 | <span style="color:red">0.426</span> |
| 27 | 2.684 | 3.146 | <span style="color:red">0.462</span> |
| 28 | 0.522 | 0.656 | 0.133 |
| 29 | 1.358 | 1.172 | 0.185 |
| 30 | 0.468 | 1.055 | 0.586 |
| 31 | 1.583 | 1.755 | 0.172 |
| 32 | 0.176 | 0.346 | 0.169 |
| 33 | 0.834 | 0.739 | 0.094 |
| 34 | 0.937 | 1.245 | 0.307 |
| 35 | 0.323 | 0.497 | 0.173 |
| 36 | 0.665 | 0.953 | 0.287 |
| 37 | 0.318 | 0.637 | 0.318 |
| 38 | 0.789 | 0.861 | 0.072 |
| 39 | 0.458 | 0.519 | 0.060 |
| 40 | 0.431 | 0.587 | 0.155 |
| 41 | 0.517 | 0.899 | <span style="color:red">0.381</span> |
| 42 | 0.621 | 0.935 | 0.313 |
| 43 | 0.184 | 0.769 | <span style="color:red">0.584</span> |
| 44 | 0.352 | 0.778 | <span style="color:red">0.426</span> |
| 45 | 0.240 | 0.580 | <span style="color:red">0.340</span> |
| 46 | 0.397 | 0.685 | 0.288 |